\documentclass[12pt]{iopart}

\usepackage{iopams} 
\usepackage{esint} 
\usepackage[dvips]{graphics, color}
\usepackage{graphicx}
\begin{document}

\title[Cavity optomechanics with Si$_3$N$_4$ membranes at cryogenic temperatures]{Cavity optomechanics with Si$_3$N$_4$ membranes at cryogenic temperatures}

\author{T. P. Purdy, R. W. Peterson, P.-L. Yu, and C. A. Regal}

\address{JILA, University of Colorado and National Institute of Standards and Technology, and
Department of Physics, University of Colorado, Boulder, Colorado 80309, USA}
\ead{tpp@jila.colorado.edu}
\begin{abstract}
We describe a  cryogenic cavity-optomechanical system that combines Si$_3$N$_4$ membranes with a mechanically-rigid Fabry-Perot cavity.  The extremely high quality-factor frequency products of the membranes allow us to cool a MHz mechanical mode to a phonon occupation of $\bar{n} < 10$, starting at a bath temperature of 5 kelvin. We show that even at cold temperatures thermally-occupied mechanical modes of the cavity elements can be a limitation, and we discuss methods to reduce these effects sufficiently to achieve ground state cooling.  This promising new platform should have versatile uses for hybrid devices and searches for radiation pressure shot noise.
\end{abstract}

\maketitle

\section{Introduction}

For decades the mechanical effects of light have been used to coax gas-phase atoms towards their quantum mechanical ground state of motion.  Recently, experimenters in the field of cavity optomechanics have learned how to extend the mechanical effects of light to more massive mesoscale objects.   A variety of microfabricated devices, with integrated optical and mechanical  or electrical and mechanical resonators, have now been employed to backaction cool mechanical objects to near their ground state:  Successful devices use lithographic electrical circuits \cite{Teufel2011,Rocheleau2010,Massel2012}, nanoscale optical-mechanical silicon resonators \cite{Chan2011}, or whispering gallery mode resonators \cite{Verhagen2012}.  These experiments combine cryogenic cooling of the mechanical object with backaction cooling of a specific mechanical mode by harnessing radiation pressure within a cavity.  

Parallel attempts have been made to cool mirrors or other dielectric objects to their ground state, and detect motion at quantum limits, within a canonical high-finesse Fabry-Perot cavity \cite{Tittonen1999a,Metzger2004,Arcizet2006b,Corbitt2007,Thompson2008,Groblacher2009b,Yang2011}.  These devices are generally characterized by lower frequency, higher quality factor mechanics, larger-mass, and power handling capabilites closer to macroscopic cavities.  Low-frequency mechanics combined with high quality factors enable longer absolute coherence times, and hence prospects for long-lived quantum memory elements.  A larger motional mass is important for searches for gravitationally induced quantum collapse \cite{Marshall2003}.   Another frontier in optomechanical systems is the realization of a quantum-limited continuous position measurement of a macroscopic object  \cite{Caves1980a,Braginsky1992,Jacobs1994,Heidmann1997,Murch2008,Yamamoto2010,Zwickl2011}.   This interest is in part motivated by large interferometers used for gravitational wave searches that will soon be limited by so-called radiation pressure shot noise \cite{Caves1980a,Braginsky1992}.  Thus far, operating optomechanical systems at high power has been one limitation to observation of radiation pressure shot noise, and Fabry-Perot cavities comprised of super mirrors and low-optical-loss mechanics are an excellent candidate for achieving the required intensities.   However, despite the promising combination of optical and mechanical properties Fabry-Perot systems afford, solid-state mechanical resonators in Fabry-Perot cavities have not yet entered the quantum regime.  One reason is that non-integrated cavities are difficult to make cryogenically compatible while maintaining optical alignment and vibrational stability. 

Fabry-Perot cavities are also particularly amenable to cooling high-tension, planar membrane mechanical objects \cite{Thompson2008}, which offer a promising route for creating hybrid systems for quantum information.  In particular, there is considerable interest in combining optomechanical and electromechanical systems to realize mechanically mediated quantum state transfer between microwave and optical photons \cite{Regal2011,SafaviNaeini2011,Taylor2011,Yu2012,Teufel2011}.  One of the successful platforms for electromechanics is a 10 MHz membrane coupled to a LC resonator, and hence one possible electro-optical transducer is a system where electrical and optical cavities are parametrically coupled to the same mechanical membrane resonator.  However, no membrane mechanical system, and in fact no mechanical object with a frequency below $\sim70$ MHz, has been brought to the quantum regime with an optical platform \cite{Chan2011,Verhagen2012}.

Here we describe the details of a robust three-component Fabry-Perot cavity at cryogenic temperatures that incorporates silicon nitride membrane microresonators \cite{Thompson2008}.  Our simple and near-monolithic design rigidly attaches the cavity mirrors and the membrane to a common base while maintaining the stringent alignment requirements of a high-finesse cavity \cite{Purdy2012}.  With cryogenic pre-cooling, we can harness the extremely high frequency quality-factor products seen in higher-order modes of Si$_3$N$_4$ membranes \cite{Zwickl2007,Wilson2009,WilsonRae2011,Yu2012}, allowing, in-principle, ground state optomechanical cooling.  By laser cooling this device in a $^4$He flow cryostat at 5 kelvin, we achieve cooling of a few MHz mechanical mode to the lowest occupations yet achieved for an oscillator with this low a frequency, $\bar{n} < 10$.  The limitation to cooling is other background mechanical modes of the cavity structure; we discuss how to understand the effects of these modes based upon the transmitted intensity spectrum and future routes to eliminating these modes.

\begin{figure} \begin{center}
\includegraphics[scale=.9]{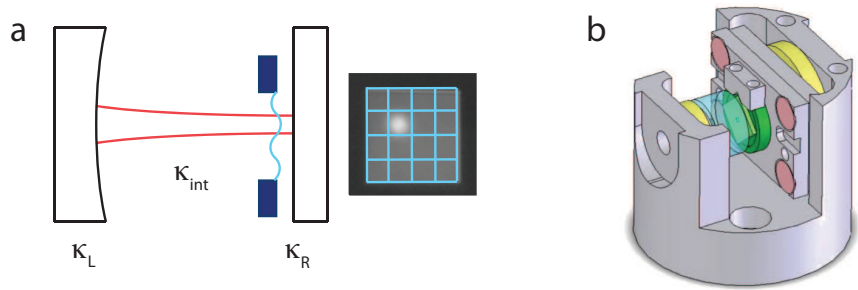} 
\caption{(a) Membrane-at-the-end cavity and simultaneous image of the intracavity optical mode and the Si$_3$N$_4$ membrane.  The nodes of the (4,4) mechanical mode are indicated in blue.  (b) Device design. Along the axial direction of the cavity from left to right are:  Piezoelectric actuator (yellow), flat cavity mirror (transparent), membrane chip, silicon holder (green) with piezoelectric actuator (yellow), and curved cavity mirror (not visible).} \label{fig1a}
\end{center}
\end{figure}

\section{Device and experiments}
Our cavity consists of two superpolished fused silica substrates with a high-reflectivity coating.  The 100 ppm throughput mirrors can have down to a few ppm scattering-absorption, creating a cavity with a finesse up to 31,000.  One mirror has a radius of curvature of 5 cm, and the other mirror is flat.  For the work presented here, the mirrors are placed $L=5.1$ mm apart and the membrane is placed 0.9 mm from the flat mirror.   This creates a ``membrane-at-the-end'' cavity as shown in Fig.~\ref{fig1a}(a).   An invar spacer connects the mirrors, the membrane, and the piezoelectric elements that translate the membrane and one mirror along the cavity axis (Fig.~\ref{fig1a}(b)). In contrast to early devices \cite{Purdy2012}, the cavity design presented here allows direct access to the membrane; by pulling the entire central metal section vertically out of the cavity, the membrane can be switched out without disturbing the high-finesse cavity mirrors.

The cavity is assembled by aligning the elements at room temperature using an optical signal and then epoxying each piece in place in turn.  We align both cavity mirrors to retroreflect a single fixed laser beam.  The membrane is then inserted into the cavity, held temporarily on a 5-axis
micrometer stage.  The transverse alignment of the membrane is assessed by in-situ imaging of both the optical mode spot, formed by a 1064 nm laser source, and the membrane, illuminated with an LED source at 940 nm where the mirror reflectivity is low (Fig.\ref{fig1a}(a)).  Any tilt in the membrane plane relative to the optical axis leads to a displacement and distortion of the optical mode spot, which also varies with the location of the membrane along the optical standing wave.  The membrane tilt is adjusted so the location and shape of the optical mode spot are not perturbed as the membrane travels over an optical wavelength.  

The cavity must be constructed in such a way to maintain alignment upon cooling the device down to cryogenic temperatures.  To this end, we employ mainly low thermal expansion materials such as invar, fused silica, and the popular cryogenic epoxy Stycast 2850FT.  Further, a key element of the construction is symmetric, uniform thickness epoxy joints that do not result in relative tilt of the elements due to differential coefficients of thermal expansion, nor in excess localized mechanical stress build-up.

The membrane inserted into the cavity is a square $d=500$ $\mu$m Si$_3$N$_4$ membrane from Norcada Inc.  We use $t=40$ nm thick membranes, as measured directly via ellipsometry.    The membrane is suspended on a 5 mm square silicon chip with a thickness of 500 $\mu$m.  We focus on cooling either the ($m$,$n$)=(2,2) or (4,4) drum modes (where $m$, $n$ are the number of antinodes in $x$, $y$) of the membrane.  These modes have resonant frequencies of 1.6 MHz and 3.2 MHz and square mode sizes of 250 $\mu$m and 125 $\mu$m respectively.   These sizes can be compared to the cavity mode, which is typically measured to have an intensity profile with a $1/e^2$ diameter of 180 $\mu$m at the membrane position.  The mechanical mode frequencies are sufficiently large to theoretically cool to $\bar{n}=0.02$ given the sideband resolution, assuming the full finesse of 31,000, even for the 1.6 MHz mode.  The silicon chip is mounted to holder also made from silicon. A compromise between high-$Q_m$ and mechanical stability is achieved by attaching the membrane chip only at three corners using Stycast 2850FT.  The mechanical damping rate is measured in-situ by observing the mechanical ringdown lifetime with an optical probe at a wavelength where the cavity has low finesse.  The realized values of $Q_m$ vary between $10^6$ and $10^7$ at 5 kelvin depending on chip mounting and membrane cleanliness.  The matched thermal expansion of the chip and holder lowers the stress on the small epoxy joints.  However, after thermal cycling from room temperature to cryogenic a few times, the epoxy joints are weakened and eventually the angular alignment of the membrane is compromised.   

\begin{figure} \begin{center}
\includegraphics[scale=1.1]{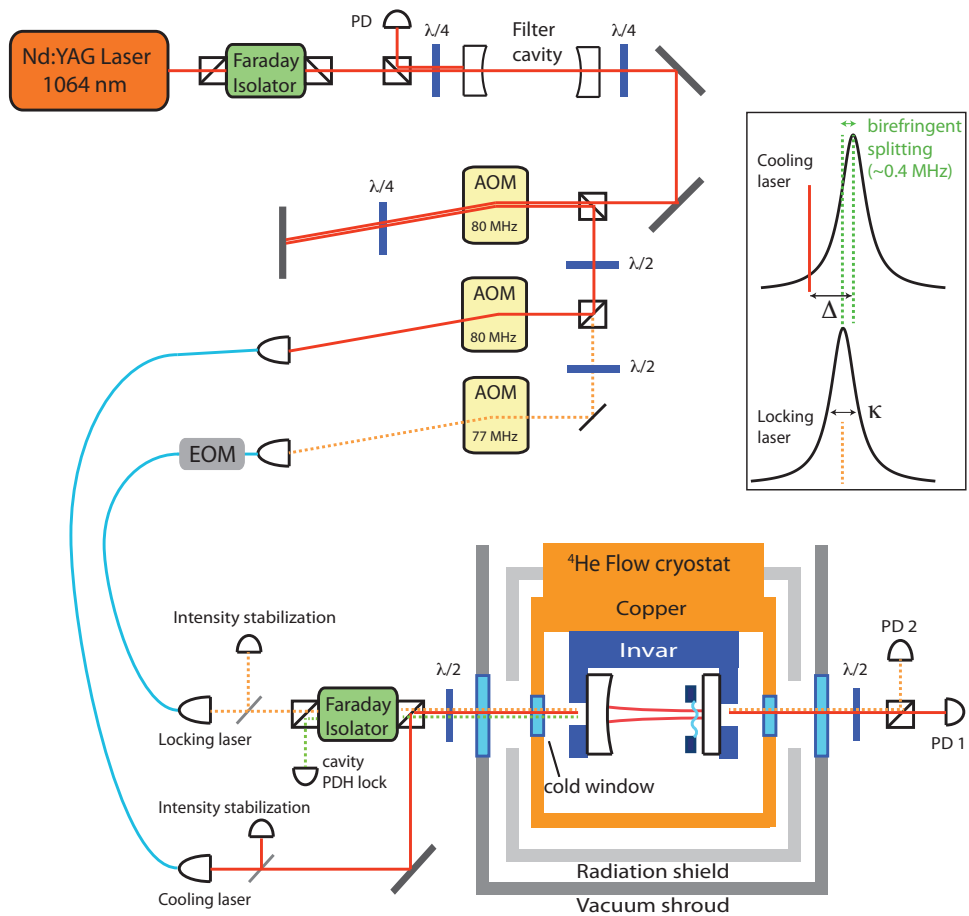} 
\caption{Experiment layout.  At the top is the laser and associated filter cavity.   The three tunable acousto-optical modulators (AOMs) are for: The high-frequency branch of the cavity-laser lock, detuning the cooling laser frequency with respect to the locking laser frequency, and setting and maintaining the intensity of each beam.  The locking laser is sent through an in-fiber electro-optical modulator (EOM) that applies frequency sidebands used for implementing a Pound-Drever-Hall lock. Multiple beams are combined and sent directly via free-space into the cryogenic cavity.} \label{fig1b}
\end{center}
\end{figure}

The entire device is cooled using a $^4$He flow cryostat from Advanced Research Systems Inc.  Additional custom radiation shielding, including cold windows, is necessary to thermalize the membrane to the temperature of the copper cold finger (Fig.~\ref{fig1b}).  Alignment is monitored upon cooling the device by assessing the finesse of the cavity and the consistency of the optical mode position as a function of the membrane translation within the cavity standing wave.   This analysis shows that we typically maintain sub-milliradian tilts between the components of the cavity.

The cooling and probing of the membrane motion uses a 1064 nm Nd:YAG laser from Innolight (Fig.~\ref{fig1b}(a)).  While already a low-noise source, we additionally pass the light through a filter cavity (40 kHz linewidth) to remove intensity and frequency noise at our MHz frequencies of interest.  Significant laser-cooling of our mechanical resonators requires careful attention to laser noise \cite{Yang2011}.  For a sense of scale, laser frequency noise of $1~\mathrm{Hz/\sqrt{Hz}}$ will result in a contribution in the output spectrum equivalent to a typical mechanical signal near the ground state.  A single tone red-detuned from the cavity resonance is used for both cooling and detection via monitoring the transmitted intensity directly on a photodetector.  We also make use of the orthogonally polarized mode for locking the relative frequency of the laser and the cavity.  At low frequencies we use the piezoelectric actuator attached to the mirror to stabilize the overall optical path length inside the cavity.  High frequency noise is eliminated by servoing the laser frequency with an AOM.  The overall bandwidth is ~100 kHz, and aggressive filtering is used to eliminate response at the membrane mechanical resonance frequencies.  Light from the two polarization modes is combined and split before and after the cavity on polarizing beam-splitters with cross-coupling of less than $10^{-3}$.  

With the membrane near the end of the cavity (Fig.~\ref{fig1a}(a)), the coupling between the membrane motion and the cavity resonance is more complicated than a membrane in the center of the cavity.  The cavity resonance frequency, linewidth, and input/output port asymmetry all depend upon the placement of the membrane within the cavity standing wave, even without absorptive loss.  These phenomena have been extensively numerically modeled in the work of Refs. \cite{Wilson2009,Wilson2011}.  A membrane placed near an end mirror can in fact have somewhat larger coupling than a membrane placed near the middle.  However, operating at positions of enhanced coupling results in degraded finesse; hence, in our experiments we actually operate at the local coupling maxima where $\kappa$ is minimized.

\section{Results}

\begin{figure} \begin{center}
\includegraphics[scale=0.44]{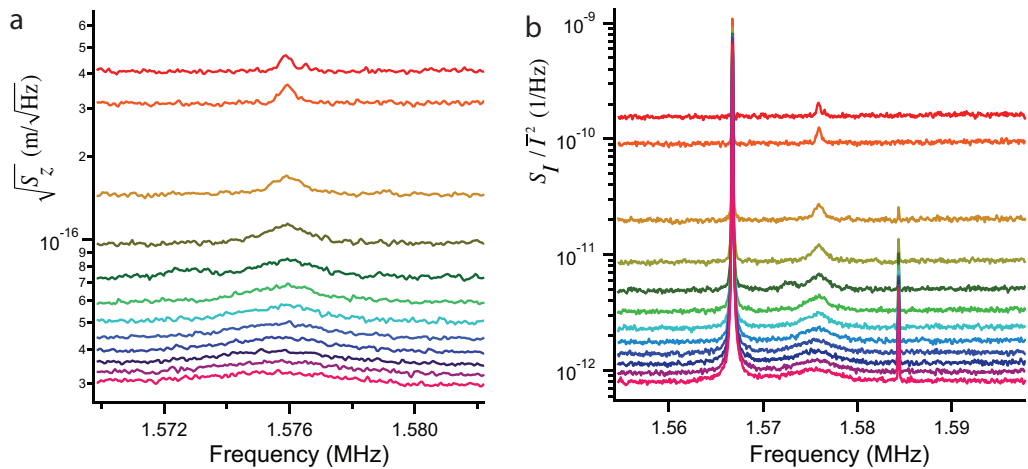} 
\caption{(a) Displacement spectrum inferred from cavity transmission near the (2,2) mechanical mode. Each curve is taken with a different cooling laser power, with cavity photon occupation ranging from $3 \times 10^5$ to $6\times10^6$ from top to bottom. For this device we operate with a cavity linewidth of $\kappa/2\pi=1.2$ MHz and a detuning of $\Delta/2\pi=-1.6$ MHz.  (b) Relative transmitted intensity spectrum corresponding to the same data as in (a).  The horizontal range has been widened to show the large peak on the left stemming from a mechanical mode of the cavity support structure.  Note, detector noise that is a dominant contributor for the low power used in this dataset is not subtracted.  Conclusions about phonon occupation for this device are displayed in Fig.~\ref{fig4}.} \label{fig2}
\end{center}
\end{figure}

In this manuscript we present results from two different cavity devices.  The first data shown in Fig.~\ref{fig2}(a) uses a device optimized for cooling the (2,2) mode.  As we increase the red-detuned input optical power, the mode is cooled and damped; the data show just the last amount of cooling.  The mechanical quality factor is the relatively large value of $Q_m=13.6\times10^6$, and hence the mode is damped from an initial linewidth $\Gamma_m=116$ mHz to an optically damped $\Gamma=5.5$ kHz (a factor of 47,000). 

\begin{figure} \begin{center}
\includegraphics[scale=0.62]{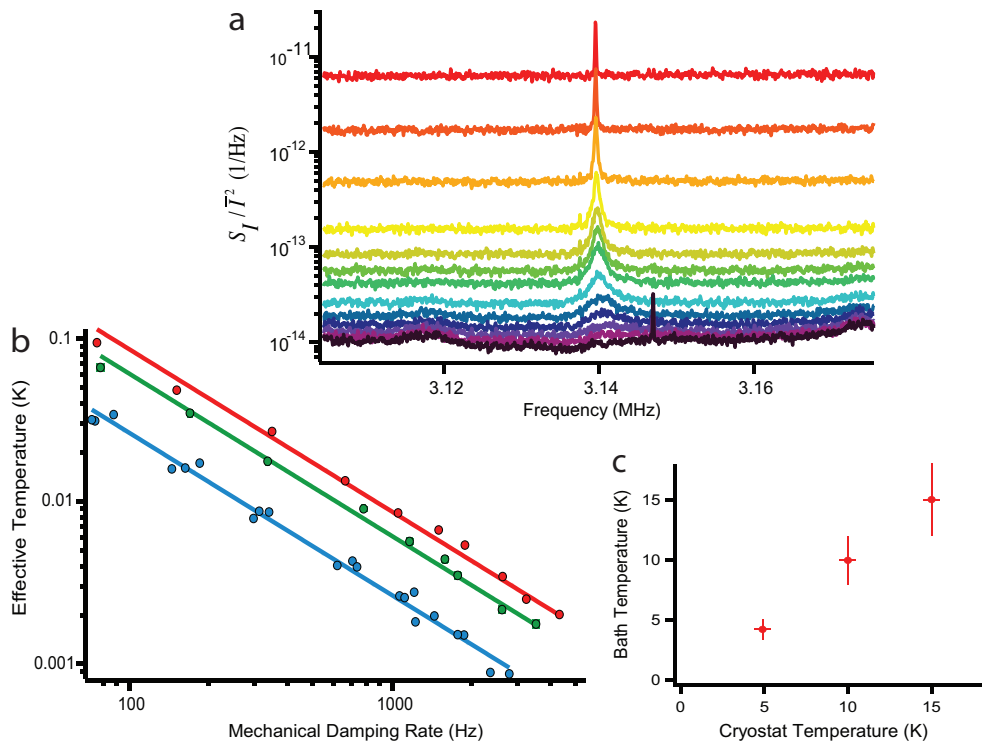} 
\caption{ (a) Cavity cooling data for the (4,4) mode device.  For this device we operate with a cavity linewidth of $\kappa/2\pi=1.4$ MHz and a detuning of $\Delta/2\pi=-2.8$ MHz.  As discussed in the text, based upon a calculation that includes the cavity noise,  for these data we reach a minimum phonon occupation of $\bar{n}=6$; note a simple conversion between the intensity spectrum and motion (Eqn.~\ref{eqn1}) yields $\bar{n}=6$ for the light blue curve (8th from the top).   Detector noise is not a dominant source here, but the detection efficiency was low due to loss in the detection path.  The small spike at 3.145 MHz in the lowest (black) dataset is electronic in origin.   (b)  Integrated mechanical response as a function of cooling power and cryostat temperature (points), along with fits to the data (lines).  The three data sets correspond to cryostat thermometer readings of 4.9 (blue), 10 (green), and 15 (red) kelvin.  The final point on the 4.9 kelvin data represents the light blue curve in (a), i.e. corresponds to $\bar{n}=6$ for a naive conversion.  (c) Bath temperature extracted from the effective temperature in (b) as a function of cryostat thermometer temperature.  The vertical error bars are the uncertainty extracted from the uncertainty in $G$ and mass, and the horizontal error bars represent the systematic uncertainty of the thermometer.} \label{fig3}
\end{center}
\end{figure}

If we expand the frequency range of the plot we see, however, that the overall relative transmitted intensity spectrum is not white (Fig.~\ref{fig2}(b)).  In these data there is a large peak to the left of the thermal peak of the (2,2) membrane mode.  We attribute this peak to a thermally occupied mode of a part of the cavity mirrors/coatings or mounting structure (which we will refer to as ``cavity mechanical modes").   Figure~\ref{fig3}(a) shows a similar set of cooling curves for a second device focusing on (4,4) mode cooling.  Here we see the cavity mechanical modes appear mainly as a weakly modulated spectrum at the highest optical powers (black curve).  Similar features are evident near the (2,2) mode if a higher laser power is used to attain a lower shot noise level.  We suspect the lower-$Q$ modes in Fig.~\ref{fig3}(a) arise from the fused silica mirror substrates based upon previous measurements in empty high-finesse cavities with similar mirrors \cite{Wilson2011,Zhao2012}, where at room temperature the modes were found to have a quality factor of 700.  Particularly high-$Q$ modes, such as the mode in Fig.~\ref{fig2} ($Q_m>20,000$) we suspect may be from the single-crystal silicon substrate holding the membrane; our simulations of the free modes of the silicon substrate indicate there are indeed coupled modes around 1.6 MHz.  In general, we have seen a variety of different cavity mechanical spectra depending on, for instance, substrate or mirror mounting techniques, coupling of the membrane motion compared to the cavity end mirrors, or the membrane mechanical quality factor.

The first step in the analysis of our data is a calibration of the membrane motion and an understanding of the membrane's bath temperature. We can extract the membrane coupling in {\it three} different ways: (1) Using thermally driven motion as a known displacement standard, (2) careful measurement of the cavity and membrane geometry including knowledge of membrane's position within the optical standing wave mode and a corresponding model of the expected coupling, and (3) the optical damping observed for a given intracavity photon number.  In the Appendix we compare these models for the (2,2) mode data and find good agreement.  The coupling for the (2,2) mode is found to be $G/2\pi=1.9\times10^{16}$ Hz/m compared to an end-mirror coupling of $\omega_c/2\pi L=5.5\times10^{16}$ Hz/m.  The (4,4) mode data presented here is not a particularly well-coupled device due to the final optical and mechanical mode overlap; for this device $G/2\pi=3.9\times10^{15}$ Hz/m.  

To verify that the membrane is thermalizing to the expected cryostat temperatures, we look at Fig.~\ref{fig3}(b) that compiles data from the (4,4) device.  We plot the integrated mechanical response from a Lorentzian fit to the data as a function of the mechanical damping (proportional to intracavity photon number) for three different cryostat temperatures.  Here we plot only the larger response data where effects of cavity mechanical mode noise can be neglected.  In Fig.~\ref{fig3}(c) we show the extracted bath temperature from these three datasets as a function of the measured cryostat temperature (taking into account slight variations in mechanical and cavity parameters with the changing bath temperature).  We find a linear trend that extrapolates to zero within the uncertainty, indicating that the membrane mechanical mode does indeed equilibrate to the cryostat cold finger temperature. 

\begin{figure} \begin{center}
\includegraphics[scale=0.6]{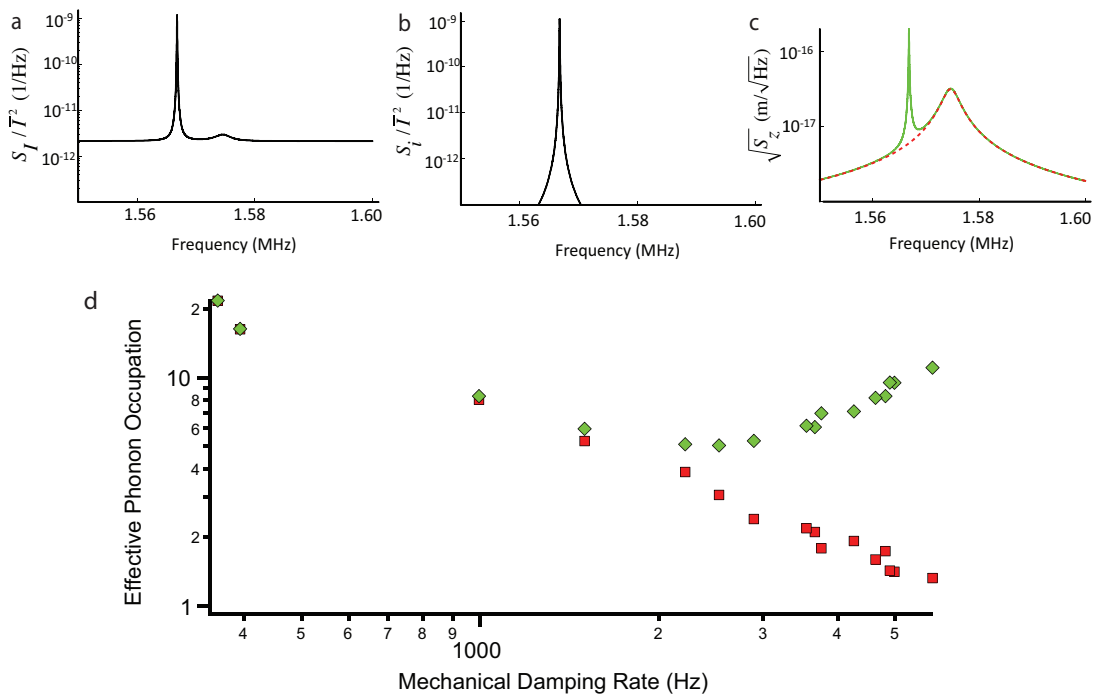} 
\caption{(a) $S_{I}(\omega)$ designed to model the observed relative intensity spectrum for the data in Fig.~\ref{fig2}(b) at an intracavity cavity photon number of $3.3\times10^6$ (teal curve 7th from the top in Fig.~\ref{fig2}(b)).  (b) Expected contributions to the relative transmission spectrum due to the cavity mechanical modes, $S_{i}(\omega)$ for the same parameters as in (a).  (c) Modeled displacement spectrum of the (2,2) mechanical mode (solid green).  The dashed red line shows the displacement spectrum expected if we were to ignore $S_{i}(\omega)$.  (d) Extracted integrated motion (converted to a phonon occupation) for the (2,2) membrane mechanical mode as a function of cooling laser power. Red squares show only the contribution of the thermally driven Lorentzian peak, which correctly predicts a small on-resonance motional spectral density, but gives an erroneously low phonon occupation.  The green diamonds include contribution from red squares and the motion induced by the cavity mechanical mode motion.  The curve in (c) corresponds to the point with a 3 kHz mechanical damping rate.} \label{fig4}
\end{center}
\end{figure}

To start we can use our extracted coupling $G$ to apply a naive conversion between the transmitted intensity spectrum and the mechanical motion of the mode of interest.  When cavity mechanical motion and radiation pressure shot noise can be neglected, we can convert the relative transmitted intensity spectrum $S_{I}(\omega)$ to a (one-sided) displacement spectrum $S_{z}(\omega)$ using the following relation: 
\begin{equation}
S_{z}(\omega)=\frac{S_{I}(\omega)}{\bar{I}^2}\frac{1}{|\Pi(\omega)|^2 G^2}.
\label{eqn1}
\end{equation}
Here $\Pi(\omega)=\chi_c(\omega)-\chi_c^*(-\omega)$
where $\chi_c(\omega)=\frac{1}{\kappa/2-\imath(\omega + \Delta)}$ is the response function for a cavity with linewidth $\kappa$ (energy decay rate) and detuning of the laser from the effective cavity resonance $\Delta$.   The extracted $S_z(\omega)$ yields a thermally driven, optomechanically damped Lorentzian on top of a floor coming from optical shot and photodetector dark noise (Fig.~\ref{fig2}(a) for example).  We extract the effective temperature of the mode by fitting the data to a Lorentzian.  We use the area under the fit to determine the mean square displacement, which is proportional to the effective temperature, and the effective phonon occupation $\bar{n}$.  Applying this procedure to the (2,2) mode data in Fig.~\ref{fig2}(a) we find the apparent effective phonon occupation displayed as the red squares in Fig.~\ref{fig4}(d), which would indicate the thermal component of the mechanical motion is brought to 1.4 quanta.

As discussed above, an important limit to laser cooling is classical noise in the relative frequency of the cooling laser and cavity resonance frequency. Such effects include laser phase noise \cite{Rabl2009b,SchleierSmith2011,Kippenberg2011} and mechanical and thermal noise in the cavity mechanical modes \cite{Pinard2000,Verhagen2012,Zhao2012}.   This relative frequency noise is converted to intensity noise in the cavity for off resonant laser drives, such as cooling tones. The intensity noise applies a radiation pressure force to the mechanical oscillator leading to an extra displacement and greater effective temperature.  Further, extracting the temperature from the transmitted intensity spectrum in the presence of classical noise becomes complex \cite{Rabl2009b,SchleierSmith2011,Yang2011}.  

As a first study of the effect of the classical noise on measurements of $\bar{n}$, we present a more thorough analysis of the data of Fig.~\ref{fig2}(b).  Here, we model the cavity mechanical mode noise as dominated by a single thermally-occupied mode.  We then determine the corresponding effect on the (2,2) mode.  Importantly, we assume that the observed cavity mechanical motion does not directly drive the membrane mode, but rather the only coupling is through the intracavity field.  If this assumption were not true, it is possible the entire output spectrum could be attributed to membrane motion.   Further, we assume the effective mass of the cavity mode is much larger than that of the membrane mode.  The physical mass of expected cavity modes is greater than $10^6$ times larger than that of membrane modes.  Hence, even without complete knowledge of the coupling to the cavity mode, this is an excellent assumption.  Thus, the cavity mechanical mode motion, to a good approximation, is unaffected by the optomechanical interaction.  In the Appendix we present a calculation for this situation.  Our model allows us to estimate the displacement spectrum of the membrane mode from the measured cavity transmission spectrum, when the simple relation between transmission and displacement spectra (Eqn. \ref{eqn1}) no longer holds.  

We estimate the functional form of the noise spectrum as a Lorentzian centered at the position of the large cavity mechanical mode peak.  Because the noise is concentrated away from the membrane mode resonance, we may assume (at least for the weakly damped data) that the area in the transmission spectrum due to the cavity mechanical mode is largely independent of the membrane motion, and is equal to $S_{i}(\omega)$, the transmission spectrum without the membrane mode of interest present.  In the Appendix it is shown that this extra intensity noise causes a radiation pressure force that drives the membrane mode to an additional displacement given by: 
\begin{equation}
S_{z,\delta f}(\omega)=\frac{4\omega_m^2 g_0^2 \bar{N}^2}{|\mathcal{N}(\omega)|^2} \frac{S_{i}(\omega)}{\bar{I}^2}Z_\mathrm{zp}^2.
\end{equation}  
In this expression, $g_0=G Z_\mathrm{zp}$ where $Z_\mathrm{zp}=\sqrt{\frac{\hbar}{2 m \omega_m}}$ is the zero point motion for an oscillator with frequency $\omega_m$ and effective mass $m$, and $\bar{N}$ is the intracavity photon number. The function $\mathcal{N}(\omega)$ represents the optomechanically modified mechanical response and is displayed in the Appendix.  

Curves displaying an example of the modeled spectrum $S_{I}(\omega)$ and the cavity mechanical noise contribution $S_{i}(\omega)$ are shown in Fig.~\ref{fig4}(a)\&(b).  The inferred value of $S_{z}(\omega)$ for these parameters is plotted in Fig.~\ref{fig4}(c) (green line).  Since the majority of the noise peak in $S_{I}(\omega)$ comes directly from $S_{i}(\omega)$,  the height of the cavity mechanical mode peak relative to the thermally driven motion in the displacement spectrum is much smaller than the corresponding ratio in the transmission spectrum.   The total estimated displacement of the membrane mode is shown in Fig.~\ref{fig4}(d) as the green diamonds.  Specifically, we plot the integrated motion $\langle z^2 \rangle$ for curves like that shown in Fig.~\ref{fig4}(c), converted to an equivalent phonon occupation.  As the laser power increases, the contribution of the cavity mechanical mode noise drive increases because the intensity fluctuations grow in proportion to the optical intensity.  Thus, a minimum phonon occupation of $\bar{n}\approx 5$ occurs at an intermediate intracavity power.  And as noted earlier, without this cavity mechanical mode (or if this mode were much farther away in frequency space) we would achieve $\bar{n}$ approaching unity, as displayed by the red squares.

Next we consider the temperature achieved for the (4,4) mode data of Fig.~\ref{fig3}(a) taking into account the cavity noise.  Here our best estimate is obtained by modeling the cavity mechanical noise spectrum as a white floor.  For these data, shot noise for the highest-power set (black curve) corresponds to a relative intensity noise of $6\times10^{-15}$ and hence we would place the classical cavity noise floor at $\sim 4\times10^{-15}$.   First, it is useful to consider the magnitude of motion this value corresponds to; as discussed above, we suspect in this case the cavity mechanical motion is dominated by the mirror fused silica substrates.  Using the $G$ values discussed above, $S_I/\bar{I}^2=4\times10^{-15}$ corresponds to membrane motion of $0.9\times10^{-17}~\mathrm{m/\sqrt{Hz}}$ or a real end-mirror motion of $0.7\times10^{-18}~\mathrm{m/\sqrt{Hz}}$.  Assuming a scaling of the mirror substrate motion with $\sqrt{T_\mathrm{bath}}$, this cavity mirror motion is within a range expected for thermally occupied modes of fused silica substrates based upon previous measurements of room temperature cavities \cite{Wilson2011}.  Finally, we apply the same model discussed above and in the Appendix to these data.  Again we assume that there is no physical coupling between the cavity mechanical modes and the membrane; since we assign the cavity mechanical modes in this situation to the mirror substrates, we believe this is a very good assumption.  We use the estimated flat spectrum at $4\times10^{-15}$ as $S_{i}(\omega)/\bar{I}^2$ and the data of Fig.~\ref{fig3}(a), and we find the phonon occupation reaches a minimum value of $\bar{n}=6$ as a function of cooling laser intensity.

In conclusion, we have presented two experimental data sets in which we have cooled Si$_3$N$_4$ membrane modes, as a conservative estimate, within a factor of 10 of the quantum mechanical ground state.  The dominant uncertainty in our temperature measurement comes from complex classical noise spectra added at the few phonon level by other mechanical modes within the cavity.  We expect with minimal additional optimization of our cavities we can remove deleterious effects of these additional cavity modes.  Already we have demonstrated the kind of system parameters required for ground state cooling; for example a fruitful set of parameters would combine the coupling and quality factors achieved for the device in Fig.~\ref{fig2} with the absence of the isolated high-$Q$ peak in Fig.~\ref{fig3}.  We believe engineering the silicon substrate to advantageously define the relative frequencies of the membrane and substrate modes will allow us to consistently achieve desired parameters in future designs \cite{Jockel2011}.  It has also been shown that relative frequency noise of the cooling laser and cavity resonance frequency can be removed via active feedback schemes involving higher-order cavity modes \cite{Zhao2012}.  Further, final studies of these devices will likely be conducted at even colder cryogenic temperatures compatible with superconducting circuits; here thermal occupation of cavity mirrors and the substrate should be reduced.

Importantly, thus far we have not observed physical absorption heating of our devices that leads to a significant increase in $T_\mathrm{bath}$.  This fact, combined with the efficient detection that these devices can afford, makes this system promising for the observation of the radiation pressure shot noise.  Even with the presence of cavity frequency noise, shot-noise from an sufficiently-strong optical tone placed on the cavity resonance should be efficiently transduced to the membrane motion. 


\ack
This work was supported by the DARPA QuASAR program, ONR YIP, and JILA NSF-PFC. We thank Konrad Lehnert's group for helpful input.  CR thanks the Clare Boothe Luce Foundation for support. TP thanks the NRC for support.

\clearpage

\appendix

\section{Calculation of thermo-mechanical noise in an optomechanical system}

	To analyze our optomechanical system we consider a standard optomechanical Hamiltonian with the addition of cavity mechanical modes of the cavity structure indexed by $i$:
\begin{eqnarray}\label{eq:Hamiltonian}
\fl	\qquad H=\hbar \omega_m c^{\dag} c + \hbar \omega_c a^{\dag} a +\hbar G Z_\mathrm{zp} (c+c^{\dag}) a^{\dag} a \\
	+ \hbar \sum_{i}{\left(\omega_i b_i^{\dag} b_i+G_i Z_{zp,i}(b_i+b_i^{\dag}) a^{\dag} a\right)} + H_{\kappa}+H_{\Gamma}  \nonumber
\end{eqnarray}
Here $c$ is the annihilation operator for the mechanical mode of interest (the membrane mode), with oscillation frequency $\omega_m$ and harmonic oscillator length $Z_\mathrm{zp}$.  $a$ is the annihilation operator for the cavity mode, at frequency $\omega_c$. $G$ is the optomechanical coupling constant, and we define a single photon coupling rate $g_0=G Z_\mathrm{zp}$.  $b_i$ is the annihilation operator for the $i$th cavity mechanical mode with frequency $\omega_i$, harmonic oscillator length $Z_{zp,i}$, and optomechanical coupling $G_i$.   Additionally, the term $H_{\kappa}$ represents the input and output optical coupling of the cavity with total cavity decay rate $\kappa=\kappa_L+\kappa_R+\kappa_\mathrm{int}$.  The contributing decay rates stem from the input port ($\kappa_L$), the output port $(\kappa_R$), and the internal loss ($\kappa_\mathrm{int}$).  The term $H_{\Gamma}$ represents the thermal drive on all of the mechanical modes.  

The optomechanical interaction can be linearized and fast oscillations at the optical frequency accounted for by defining $a(t)=(\bar{a}+d(t))e^{-\imath \omega_L t}$.  $\bar{a}=\frac{\sqrt{\kappa_L} \bar{a}_\mathrm{in}}{\kappa/2-\imath \Delta}$ is the large classical amplitude of the intracavity field and $d(t)$ represents small fluctuations about this value.  $\bar{a}_\mathrm{in}$ is the coherent state amplitude of the input laser field driven at frequency $\omega_L$.

   From the Hamiltonian in Eqn.~(\ref{eq:Hamiltonian}), we derive a set of Heisenberg-Langevin equations of motion and transform them into the frequency domain using the Fourier transformation $f(\omega)=\int^{\infty}_{-\infty}e^{\imath \omega t} f(t) dt$, $f^{\dag}(\omega)=\int_{-\infty}^{\infty}e^{\imath \omega t} f^{\dag}(t) dt$.
\[
\fl  d(\omega)=\chi_c(\omega) \left(-\imath \bar{a} G z(\omega)  -\imath \bar{a} \sum{G_i z_i(\omega)}+ \sqrt{\kappa_L}\xi_L(\omega) +\sqrt{\kappa_\mathrm{int}}\xi_\mathrm{int}(\omega)+\sqrt{\kappa_R} \xi_R(\omega)\right)
\]
 \[
\fl  \frac{z(\omega)}{Z_\mathrm{zp}}=\imath g_0 \Big(\chi_m^*(-\omega)-\chi_m(\omega) \Big)\Big(\bar{a} d^{\dag}(\omega) +\bar{a}^*d(\omega)\Big)+\sqrt{\Gamma_m}\Big(\chi_m(\omega)\eta(\omega)+\chi_m^*(-\omega)\eta^{\dag}(\omega)\Big)
\]
\[
\fl \frac{z_i(\omega)}{Z_{zp,i}}=\sqrt{\Gamma_i}\left(\chi_{i}(\omega)\eta_i(\omega)+\chi_{i}^*(-\omega)\eta_i^{\dag}(\omega) \right)
\]
$z$ and $z_i$ represent the small displacements of the mechanical modes about their optomechanically shifted equilibrium positions $\bar{z}$ and $\bar{z_i}$, such that $Z_\mathrm{zp}\left(c+c^{\dag}\right)=\bar{z}+z$ and $Z_{zp,i}\left(b_i+b_i^{\dag}\right)=\bar{z}_i + z_i$.  We use here the mechanical susceptibilities $\chi_{m}(\omega)=\frac{1}{\Gamma_m/2-\imath(\omega-\omega_{m})}$ and $\chi_{i}(\omega)=\frac{1}{\Gamma_i/2-\imath(\omega-\omega_{i})}$.  $\chi_c(\omega)=\frac{1}{\kappa/2-\imath(\Delta+\omega)}$ is the cavity susceptibility, where $\Delta=\omega_c-\omega_L-G \bar{z}-\sum{G_i \bar{z}_i}$ is the detuning of the laser input frequency from optomechanically shifted cavity resonance frequency.  The operators $\xi_L e^{-\imath \omega_L t}$,  $\xi_R e^{-\imath \omega_L t}$, and $\xi_\mathrm{int} e^{-\imath \omega_L t}$ are Langevin noise operators representing vacuum fluctuations entering the cavity from the input, loss, and output ports (see for example \cite{WallsandMilburn,Clerk2010}).  $\eta$ and $\eta_i$ are the Langevin noise operators representing the thermal and vacuum noise driving the mechanical modes. To simplify the equations of motion we drop small terms of order $d^2$, $d z$, and $d z_i$.  Additionally, we work in the limit where $G_i^2 Z_{zp,i}^2 \bar{a}^*\bar{a}/\kappa \ll \bar{n}_{th,i} \Gamma_i$ where $\bar{n}_{th,i}=\frac{k_b T_\mathrm{bath}}{\hbar \omega_i}$ and $T_\mathrm{bath}$ is the bath temperature of the modes.  In this limit the optical drive does not perturb the mechanical state of the cavity mechanical modes.  However, signatures of the cavity mechanical modes are still imprinted on the optical mode.
	
	We can then solve the equations of motion and calculate the mechanical displacement spectrum $S_z^{(2)}(\omega)=\left<z(-\omega) z(\omega)\right>$.
 \begin{eqnarray}\label{eq:zz}
\fl	\qquad \frac{\left<z(-\omega) z(\omega)\right>}{Z_\mathrm{zp}^2}= \frac{1}{\mathcal{N}(-\omega) \mathcal{N}(\omega)}\bigg\{&
	  \Gamma_m \left(\frac{\bar{n}_\mathrm{th}+1}{|\chi_m(\omega)|^2}+\frac{\bar{n}_\mathrm{th}}{|\chi_m(-\omega)|^2}\right) \\
	 &+ 4\omega_m^2 g_0^2 \kappa \bar{a}^*\bar{a}|\chi_c(-\omega)|^2 \nonumber \\
	& +4\omega_m^2 g_0^2(\bar{a}^*\bar{a})^2 |\Pi(\omega)|^2 \left< \delta f (-\omega) \delta f(\omega)\right> \bigg\} \nonumber
\end{eqnarray}
where $\mathcal{N}(\omega)=\frac{1}{\chi_m(\omega)\chi_m^*(-\omega)} -\imath 2 \omega_m g_0^2 \bar{a}^*\bar{a} \Pi(\omega) $ and $\Pi(\omega)=\chi_c(\omega)-\chi_c^*(-\omega)$.  We make use of the operator expectation values $\left<\xi_L(-\omega)\xi_L^{\dag}(\omega)\right>=\left<\xi_\mathrm{int}(-\omega)\xi_\mathrm{int}^{\dag}(\omega)\right>=\left<\xi_R(-\omega)\xi_R^{\dag}(\omega)\right>=1$, and $\left<\eta(-\omega) \eta^{\dag}(\omega)\right>=\bar{n}_\mathrm{th}+1$, $\left<\eta^{\dag}(-\omega) \eta(\omega)\right>=\bar{n}_\mathrm{th}$.  The first term of Eqn.~(\ref{eq:zz}) contains the thermal motion of the membrane.  The second term includes the membrane motion induced by radiation pressure from optical shot noise.  The effect of the cavity mechanical modes is seen in the third term, through $\left< \delta f (-\omega) \delta f(\omega)\right>= \sum{G_i^2 \left< z_i(-\omega) z_i(\omega)\right>}$, which represents the noise spectrum of the cavity frequency shifts induced by the cavity mechanical modes.   

	We next compute the spectrum of intensity fluctuations for light directly detected on a photodetector at the output port of the cavity.  Let $S_{II}^{(2)}=\left<\left(I(-\omega)-\bar{I}\right)\left(I(\omega)-\bar{I}\right) \right>$ be the two-sided power spectrum of the detected photocurrent where $I(t)=\epsilon\hbar \omega_L \mathcal{R} a^{\dag}_\mathrm{out}(t) a_\mathrm{out}(t)+(1-\epsilon) a_n^{\dag}(t)a_n(t)+I_d(t)$, with mean value $\bar{I}=\left<I\right>$.  $\mathcal{R}=\frac{q_e}{\hbar \omega_L}$ is the photodetector sensitivity where $q_e$ is the electron charge. $\epsilon$ is the detection efficiency, $a_n(t)=\xi_n(t)e^{-\imath \omega_L t}$ is a Langevin noise operator representing vacuum fluctuations entering the detector through the loss port associated with the detector inefficiency, and $I_d$ is the photodetector dark current.  The output optical field $a_\mathrm{out}(t)=(\bar{a}_\mathrm{out}+d_\mathrm{out}(t))e^{-\imath \omega_L t}$ is evaluated via the input-output relations: $\bar{a}_\mathrm{out}=\sqrt{\kappa_R}\bar{a}$ and $d_\mathrm{out}=\xi_R+\sqrt{\kappa_R} d$.
\begin{eqnarray}
\fl	\qquad \frac{S_{II}^{(2)}(\omega)}{\bar{I}^2}=\frac{1}{\kappa_R (\bar{a}^*\bar{a})^2}\Big< &\left[ \bar{a}^*(\sqrt{\kappa_R} d(-\omega)-\xi_R(-\omega))+\bar{a} (\sqrt{\kappa_R}d^{\dag}(-\omega)-\xi_R^{\dag}(-\omega))\right]\nonumber \\
	  & \times   \left[ \bar{a}^*(\sqrt{\kappa_R} d(\omega)-\xi_R(\omega))+\bar{a}(\sqrt{\kappa_R}d^{\dag}(\omega)-\xi_R^{\dag}(\omega))\right]  \Big>\nonumber\\
	 &  +  \frac{1-\epsilon}{\epsilon}\frac{1}{\kappa_R \bar{a}^*\bar{a}}+\frac{\left<I_d(-\omega) I_d(\omega)\right>}{\bar{I}^2}\nonumber\\
	&  =  \left<\Psi(-\omega) \Psi(\omega)\right> + \frac{1-\epsilon}{\epsilon}\frac{1}{\kappa_R \bar{a}^*\bar{a}}+\frac{\left<I_d(-\omega) I_d(\omega)\right>}{\bar{I}^2}\nonumber
\end{eqnarray}
where $\Psi(\omega)=\Psi_q(\omega)+\Psi_m(\omega)+\Psi_i(\omega)$ contains the following components
\begin{eqnarray}
\fl	\qquad \Psi_{q}(\omega)=\frac{1}{\bar{a}^*\bar{a}}&\bigg(\bar{a}^*\sqrt{\kappa_L}\chi_c(\omega) \xi_L(\omega)+\bar{a}\sqrt{\kappa_L}\chi_c^*(-\omega) \xi_L^{\dag}(\omega)+\frac{\bar{a}^*}{\sqrt{\kappa_R}}(\kappa_R \chi_c(\omega)-1)\xi_R(\omega)\nonumber\\
  	+ & \frac{\bar{a}}{\sqrt{\kappa_R}}(\kappa_R \chi_c^*(-\omega)-1)\xi_R^{\dag}(\omega)+\bar{a}^*\sqrt{\kappa_\mathrm{int}}\chi_c(\omega) \xi_\mathrm{int}(\omega)+\bar{a}\sqrt{\kappa_\mathrm{int}}\chi_c^*(-\omega) \xi_\mathrm{int}^{\dag}(\omega)\bigg)\nonumber
\end{eqnarray}
\begin{eqnarray}
\fl	\qquad	\Psi_{m}=-\imath G \Pi(\omega) z(\omega)\nonumber
\end{eqnarray}
\begin{eqnarray}
\fl	\qquad	\Psi_{i}=-\imath \Pi(\omega)  \delta f(\omega).\nonumber
\end{eqnarray}
We can thus compute the functions
\begin{eqnarray}
\fl	\qquad		\left<\Psi_{q}(-\omega) \Psi_{q}(\omega)  \right>=\frac{1}{\kappa_R \bar{a}^*\bar{a}}\nonumber
\end{eqnarray}
\begin{eqnarray}
\fl	\qquad	\left<\Psi_{m}(-\omega) \Psi_{m}(\omega)  \right>= G^2|\Pi(\omega)|^2 \left<z(-\omega) z(\omega)\right>\nonumber
\end{eqnarray}
\begin{eqnarray}
\fl	\qquad	\left<\Psi_{i}(-\omega) \Psi_{i}(\omega)  \right>=|\Pi(\omega)|^2 \left< \delta f (-\omega) \delta f(\omega)\right>\nonumber
\end{eqnarray}
and the nonzero cross terms are
\[
\fl \qquad \left<\Psi_q(-\omega) \Psi_m(\omega) \right>+\left<\Psi_m(-\omega) \Psi_q(\omega) \right>=-4 \omega_m g_0^2 \mathrm{Im} \left[\frac{\Pi(\omega)}{\mathcal{N}(\omega)} \chi_c(-\omega)  \right]
\]
\[
\fl \qquad \left<\Psi_i(-\omega) \Psi_m(\omega) \right>+\left<\Psi_m(-\omega) \Psi_i(\omega) \right>=-4 \omega_m \bar{a}^*\bar{a} g_0^2 \mathrm{Im}\left[\frac{\Pi(\omega)}{\mathcal{N}(\omega)}  \right] |\Pi(\omega)|^2 \left< \delta f (-\omega) \delta f(\omega)\right>.
\]
The photodetector signals we record in the experiment are one-sided power spectra $S_{I}(\omega)=S_{II}^{(2)}(-\omega)+S_{II}^{(2)}(\omega)$.  Similarly we define $S_{i}(\omega)/\bar{I}^2=2 	\left<\Psi_{i}(-\omega) \Psi_{i}(\omega)  \right>$, the one-sided spectrum of transmitted intensity noise due to the fluctuations of the cavity mechanical modes, and from this we infer $S_{z}(\omega)=\left<z(-\omega) z(\omega)\right>+	\left<z(\omega) z(-\omega)\right>$ the one-sided displacement spectrum.

	If the contribution of the cavity mechanical modes is small ($S_i(\omega) \rightarrow 0 $) and the radiation pressure shot noise is small, then the displacement spectrum can be easily inferred from the transmission spectrum.
\[
	S_{z,\delta f \rightarrow 0}(\omega)+ S_{\mathrm{noise floor}}(\omega) = \frac{1 }{G^2 |\Pi(\omega)|^2} \frac{S_{I}(\omega)}{\bar{I}^2}
\]
Where $S_{\mathrm{noise floor}}(\omega)=\frac{2}{G^2 |\Pi(\omega)|^2}\left( \frac{1}{\epsilon}\frac{1}{\kappa_R \bar{a}^* \bar{a}}+\left<I_d(-\omega) I_d(\omega)\right>  /\bar{I}^2 \right)$ is the detection noise floor due to optical shot noise and detector noise.  If frequency noise is not negligible and $S_{i}(\omega)$ can be measured empirically then the contribution to $S_{z}(\omega)$ from the optomechanically transduced fluctuations of the cavity mechanical modes can be computed from Eqn.~(\ref{eq:zz}).  
\begin{eqnarray}
	\frac{S_{z,\delta f}(\omega)}{Z_\mathrm{zp}^2}=\frac{4\omega_m^2 g_0^2 (\bar{a}^*\bar{a})^2}{|\mathcal{N}(\omega)|^2} \frac{S_{i}(\omega)}{\bar{I}^2} \label{Eqn:final}
\end{eqnarray}
 	Note $\bar{a}^*\bar{a}$ is the intracavity photon number $\bar{N}$. From $S_{z}(\omega)$ the root mean square displacement of the membrane mode can be computed.
\[
		\frac{\left<z^2\right>}{Z_\mathrm{zp}^2}=\int_0^{\infty} \frac{S_{z}(\omega)}{Z_\mathrm{zp}^2} \frac{d\omega}{2 \pi}=2 \left( \bar{n}+\frac{1}{2}\right)
\]

We can compare our final expression for the impact of mechanical modes within the cavity (Eqn.~\ref{Eqn:final}) to the result expected for equivalent frequency noise on a laser at the input port of the cavity.  We find our expression can be translated into a formula equivalent to the laser frequency-noise result derived in, for example, Ref. \cite{SchleierSmith2011}.

\section{Calibration of membrane motion}

Here we present the three different methods we use to calibrate the mechanical motion and corresponding temperature.  We compare the methods by assessing the membrane coupling extracted in each case for the (2,2) data presented in Fig.~\ref{fig2}; based on these analyses we deduce an uncertainty in $G$ of 5$\%$.

\subsection{Bath temperature and optical damping}

The thermal motion of the mechanical resonator is equated to the effective temperature $T$ via $\langle z^2 \rangle=\frac{k_b T}{m \omega_m^2}$.   Optomechanical cooling theory in the large damping limit tells us $T=T_\mathrm{bath}\frac{\Gamma_m}{\Gamma}$, where $\Gamma$ can simply be determined by the measured linewidth of the optically damped mechanical resonator.  We compare $\langle z^2 \rangle$ as determined via integrating the spectrum $\int_0^\infty S_z(\omega) \frac{d\omega}{2\pi}$ to that from $\frac{k_b T_\mathrm{bath}}{m \omega_m^2}\frac{\Gamma_m}{\Gamma}$.  $T_\mathrm{bath}=4.9$ K is input based upon the cryostat thermometer, where the accuracy of this measurement was motivated via the trend in Fig.~\ref{fig3}(c).  $S_z(\omega)$ can be extracted from the measured intensity spectrum via Eqn.~\ref{eqn1} with one free parameter $G$.

The inputs to this calculation are:  $\kappa/2\pi = 1.2$ MHz, which is determined by a measurement of the cavity linewidth in ringdown for the position of the membrane during the measurement.  $\Delta/2\pi = 1.6$ MHz, which is determined from the detuning set with respect to the on-resonant locking light, accounting for a birefringent splitting of 0.4 MHz.  The mechanical frequency $\omega_m/2\pi = 1.575$ MHz.  The mechanical quality factor $Q_m=13.6\times10^6$.  The effective mass $m=\rho d^2 t/4$, which is determined using $\rho=2700$ $\mathrm{kg/m^3}$ \cite{Verbridge2006}.  (However, given the range of values in the literature for LPCVD Si$_3$N$_4$ \cite{Wilson2011} we would place a $10\%$ systematic uncertainty on the mass.  Note, the mass uncertainty is relevant for $G$ but not $g_0$ \cite{Gorodetsky2010}.)  

This comparison tells us $G/2\pi =1.8 \times 10^{16}$ Hz/m.

\subsection{Membrane-at-the-end model}

With knowledge of the position of the membrane within the cavity standing wave we can model the expected coupling.  In each iteration of the experiment, we scan the membrane within the cavity to sit at the position for which the cavity linewidth $\kappa$ is minimized \cite{Wilson2011}.  With knowledge of the cavity and membrane parameters we can calculate the expected coupling at this operating position we refer to as $Z_\mathrm{min}$.  $Z_\mathrm{min}$ is registered within the standing wave $\sim0.9$ mm from the flat mirror of the cavity.  The cavity has an overall length of 5.1 mm, which is measured via the ratio of the transverse mode spacing to the free spectral range of 29.4 GHz.  Based on direct ellipsometry measurements of our film we base our calculation on a $t=40$ nm thick membrane with index $n=2.0$.   Hence, at $Z_\mathrm{min}$ we predict $\mathrm{d}\omega_c/\mathrm{d}z=2\pi \times 2.9\times10^{16}$ Hz/m. 

We then apply a correction based upon the measured transverse mode overlap between the mechanical mode of interest and the $\mathrm{TEM}_{00}$ cavity mode (an example of such a measurement is shown in Fig.~\ref{fig1a}).  To account for the mode overlap we calculate $G=\eta \frac{d\omega_c}{dz}$ using \cite{Gillespie}
\begin{equation}
\eta_{mn} = \left | \iint \mathrm{d}x\mathrm{d}y I(x,y)a_{mn}(x,y)/a_0 \right |
\end{equation}
where $a_{mn}=a_0 \sin(m\pi x / d) \sin(n\pi y /d)$ and $I(x,y)$ is a normalized intensity function $I(x,y)=\frac{2}{\pi \mathrm{w}_x \mathrm{w}_y}\mathrm{exp}(-2(x-x_0)^2/\mathrm{w}_x^2)\mathrm{exp}(-2(y-y_0)^2/\mathrm{w}_y^2)$.  We measure the location of the (2,2) mode while cold to be at the coordinate position $(x_0,y_0)=(108, 99)$ $\mu$m.  The size of the mode in $x$,$y$ is measured to be $\mathrm{w}_x=92$ $\mu$m and $\mathrm{w}_y=88$ $\mu$m.  Hence $\eta=0.67$, and $G/2\pi = 2.0 \times 10^{16}$ Hz/m.

\subsection{Optical damping and intracavity photon number}

The optical damping $\Gamma$ we observe for a given intracavity photon number is also a measure of $G$.  To infer the intracavity photon number $\bar{N}$ from the measured output flux of the cavity we understand the asymmetry of the cavity and the internal loss.  Usually we orient the cavity according to Fig.~\ref{fig1a} where the membrane is at the output (right) side of the cavity, providing the most signal at the output.  However, for these particular measurements we happened to measure out the less-transmissive port of the cavity, i.e. the membrane and flat mirror were actually at the input (left) side of the cavity.  For a measurement of the photocurrent $\bar{I}$ at the output port the photon number is given by 
\begin{equation}
\bar{N}=\frac{P_\mathrm{out}}{\hbar \omega_c \kappa}\frac{\kappa}{\kappa_R}=\frac{1}{\kappa_R}\frac{\bar{I}}{q_e}\frac{1}{\epsilon}
\end{equation}
for a low-reflectivity membrane \cite{Wilson2011}, where $\epsilon=\epsilon_d \epsilon_p$ is a combination of two efficiency factors.  In our setup for these measurements, the detector efficiency is $\epsilon_d=0.87$ and the propagation losses from the cavity output to the detector are given by $\epsilon_p=0.88$.  

To find $\kappa_R$ we must understand all the contributions to $\kappa=\kappa_R+\kappa_L+\kappa_\mathrm{int}$.  $\kappa_\mathrm{int}$ is dominated by clipping of the transverse mode on the silicon frame that results from imperfect alignment when cooling to cryogenic temperatures, and hence for modeling $G$ and $\kappa$ we consider this a loss that is independent of membrane position.  At the position that corresponds to the minimum value of $\kappa$ ($Z_\mathrm{min}$), the theoretical analysis of the three-element cavity described above indicates $\kappa_\mathrm{min}=0.79$ MHz.  With a well-aligned cavity we often achieve this value; in this particular cavity at room temperature we achieve $\kappa_\mathrm{min}=0.85$ MHz, and when cooled down we find $\kappa_\mathrm{min}=1.17$ MHz.  This indicates an internal loss contribution of $\kappa_\mathrm{int}=0.33 \kappa$.  The ratio between $\kappa_R$ and $\kappa_L$ can be calculated from the asymmetry of the cavity, which can be determined from a calculation of the resonant reflection $R$ and transmission $T$ at our operating position of $Z_\mathrm{min}$. We use the expression $\kappa_L/\kappa_R=\frac{(1+\sqrt{R})^2}{T}$ to find $\kappa_L/\kappa_R=1.9$  \cite{Wilson2011}.  Thus $\kappa_R=0.23 \kappa$ taking into account $\kappa_\mathrm{int}$.

We then compare the measured optically damped linewidth $\Gamma$ to the expected calculated $\Gamma$ given by \cite{Marquardt2007a,WilsonRae2007} 
\begin{equation}
\Gamma = G^2Z_\mathrm{zp}^2\kappa \bar{N} \left(|\chi_c(\omega_m)|^2-|\chi_c(-\omega_m)|^2 \right).
\end{equation} 
This gives us $G/2\pi =1.9 \times 10^{16}$ Hz/m.

\section*{References}


\end{document}